\documentclass[twocolumn,useAMS,usenatbib]{mn2e}

\usepackage{color}
\usepackage{amsmath}
\usepackage{graphicx}
\usepackage{amssymb}
\usepackage{psfrag}

\newcommand{\beq}{\begin{equation}}
\newcommand{\eeq}{\end{equation}}
\newcommand{\bay}{\begin{array}}
\newcommand{\eay}{\end{array}}
\newcommand{\beqa}{\begin{align}}
\newcommand{\eeqa}{\end{align}}
\renewcommand{\eeqa}{\end{align}}

\newcommand{\nn}{\nonumber}
\newcommand{\bB}{{\bf{B}}}

\newcommand{\brac}[1]{\left({#1}\right)}

\newcommand{\pd}[2]{\frac{\partial{#1}}{\partial{#2}}}

\newcommand{\curl}{\nabla\times}
\renewcommand{\div}{\nabla\cdot}

\newcommand{\fmag}{\boldsymbol{\mathfrak{F}}_{\textrm{mag}}}

\newcommand{\clM}{\mathcal{M}}

\newcommand{\rmp}{\mathrm{p}}
\newcommand{\rhop}{\rho_\rmp}
\newcommand{\red}[1]{{\color{black}{#1}}}

\title[Magnetically-driven NS crustquakes]
         {Magnetically-driven crustquakes in neutron stars}
\author[S. K. Lander et al.]
       {S. K. Lander${}^{1,2}$\thanks{skl@soton.ac.uk}, N. Andersson${}^2$,
         D. Antonopoulou${}^1$ and A. L. Watts${}^1$\\ \\
${}^1$ Astronomical Institute ``Anton Pannekoek'', University of
Amsterdam, Postbus 94249, 1090 GE Amsterdam, the Netherlands\\
${}^2$ Mathematical Sciences, University of Southampton, Southampton
SO17 1BJ, UK}

\begin{document}

\pagerange{\pageref{firstpage}--\pageref{lastpage}} \pubyear{0000}
\maketitle

\label{firstpage}

\begin{abstract}
Crustquake events may be connected with both
rapid spin-up `glitches' within the regular slowdown of neutron stars,
and high-energy magnetar flares. We argue that magnetic field decay builds up stresses in a
neutron star's crust, as the elastic shear force resists the Lorentz
force's desire to rearrange the global magnetic-field
equilibrium. We derive a criterion for crust-breaking induced by a changing
magnetic-field configuration, and use this to investigate strain patterns in a
neutron star's crust for a variety of different magnetic-field
models. Universally, we find that the crust is most liable to break if
the magnetic field has a strong toroidal component, in which case the epicentre of
the crustquake is around the equator. We calculate the energy
released in a crustquake as a function of the fracture depth, finding that it
is independent of field strength. Crust-breaking is, however, associated with a characteristic
local field strength of $2.4\times 10^{14}$ G for a breaking strain of
$0.001$, or $2.4\times 10^{15}$ G at a breaking strain of $0.1$. We find
that even the most luminous magnetar giant flare could have been
powered by crustal energy release alone.
\end{abstract}

\begin{keywords}
\red{stars: neutron -- stars: magnetic fields -- stars: magnetars -- asteroseismology}
\end{keywords}

\section{Introduction}

The crust of a neutron star (NS) is a rigid elastic shell around a
kilometre thick, which connects the supranuclear-density fluid core
with the star's magnetosphere, and in turn any observable
phenomena. As for any elastic medium, however, there is a maximum
strain it can sustain -- beyond which the crust will yield locally,
causing seismic activity or `crustquakes'.

A crustquake scenario related to changes in rotational strain was
suggested shortly after the discovery of radio pulsars, as a way to
explain observations that the otherwise stable spindown of a pulsar
can be interrupted by abrupt increases -- `glitches' -- in spin frequency and spin-down
rate \citep{baym_ppr}. The idea is that the rotational oblateness at
the star's birth is frozen into the crust; as the star spins down it
wants to become more spherical, and the overly-oblate crust develops strains
which eventually break it and cause an increase in angular momentum of
the crust. This mechanism alone, however, cannot
explain the observed timing behaviour; instead, in the currently
standard glitch scenario the spin-up is attributed to a sudden
transfer of angular momentum from a more rapidly rotating superfluid
component to the rest of the star \citep{and_itoh}. Nonetheless, crustquakes are often invoked in
glitch models, either as a trigger for these sudden angular-momentum
transfer events \citep{le96,Eich10} or to explain the persistent
changes in spin-down rate seen after some glitches \citep{accp94}.

\begin{figure*}
\begin{center}
\begin{minipage}[l]{0.8\linewidth}
\psfrag{decay}{{\bf field decay/rearrangement}}
\includegraphics[width=\linewidth]{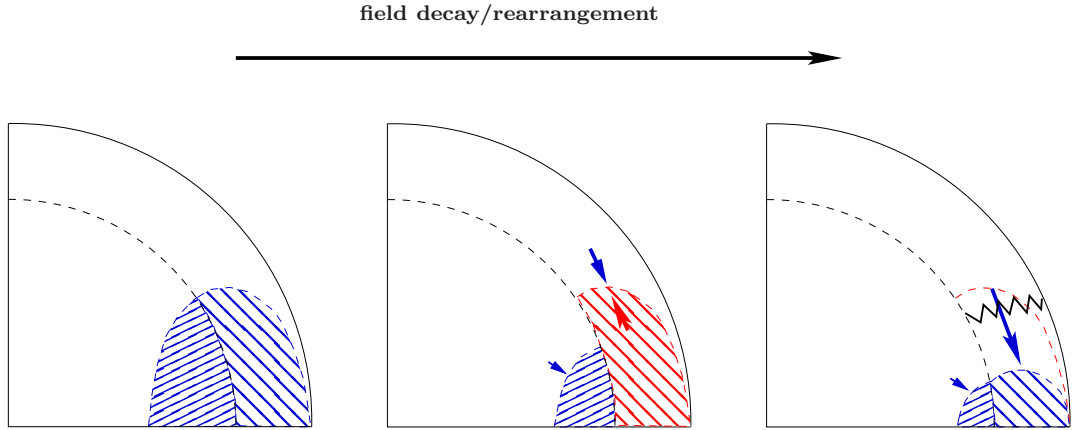}
\end{minipage}
\caption{\label{crust-crack}
               Cartoon of crust-breaking scenario. For clarity we have shown the motion of an equatorial region of magnetic
               flux to represent field rearrangement, but the argument is applicable to any local changes in the field
               anywhere in the crust. We assume the young NS has reached a
               hydromagnetic equilibrium by the time the crust freezes, so
               that the crust does not initially need to support any
               stresses (left-hand plot). At some later point in the star's evolution the magnetic field has lost
               energy and would need to adjust to remain in a global
               fluid equilibrium, but whilst this adjustment may take
               place in the fluid core, it is resisted by shear
               stresses in the crust (middle plot). The magnetically-induced stresses to the crust build, and
               eventually some region of the crust will exceed its
               yield strain and break (right-hand plot). The local magnetic field will
               be able to return to a fluid equilibrium again.}
\end{center}
\end{figure*}

In addition to these rotational effects, magnetic stresses will also develop in the crust throughout a NS's
lifetime, as a result of internal magnetic field evolution.  
For typical radio pulsars such stresses might be negligible, since the
crust's elastic energy exceeds the magnetic energy, and the elastic
force dominates the Lorentz force. For highly magnetised NSs
however, like magnetars -- objects with inferred dipole magnetic
fields at least as high as $\sim\!10^{15}$ G -- the two energies are
comparable, and it is quite feasible that magnetic stresses could be
strong enough to induce crust-yielding: crustquakes or plastic
flow. Such magnetically-driven seismic activity forms the core of the widely-accepted model for magnetar
activity, firstly put forward by \citet{thom_dunc95}
to explain bursts in anomalous X-ray pulsars (AXPs) and the bursts and
gamma-ray giant flares in soft-gamma ray repeaters (SGRs). The
recurrent bursts in magnetars have characteristic durations in the range
$\sim\!0.01-1\ \rm{s}$ and peak luminosities up to $10^{41}\ \rm{erg\ s}^{-1}$,
and are in many cases associated with glitches or other timing
anomalies \citep{wt06,dib_kaspi14}. 
The potential connection with crustquakes is consistent with the
observation that the burst-energy distribution in magnetars follows a
power law \citep{cheng96,gogus00}, similar to that of earthquakes.

Recent observations are indicative of a continuum of activity in radio
pulsars and magnetars \citep{kaspi}: SGRs have been discovered with weak inferred dipole
fields (see, e.g. \citet{rea10}), and magnetar-like activity has been
seen from some (otherwise rotationally-powered) radio pulsars, such as
the burst and coincident glitch in J1846-0258 \citep{ggg+08,kh09}.
This has led to considerable efforts to explain the different
phenomenologies of NSs in a unified scheme by studying the
thermal and magnetic-field evolution in their crusts \citep{perna_pons,PonsRea12}.
These first results suggest that seismic activity induced by
magnetic field evolution is of relevance not only for magnetars, but
also rotationally-powered pulsars.

Motivated by the many possible observational manifestations of
crustal stresses in a NS, we study a mechanism in
which a rearranging global magnetic field provides the source of these
stresses, eventually causing the crust to yield. We
derive a condition for magnetically-induced crustal failure based on
the von Mises criterion for the yielding of elastic media. Using a
variety of different magnetic-field models, including NSs with normal and
superconducting cores and with a force-free magnetosphere, we study
the crustal strain patterns that these field configurations would
produce and the point at which regions of the crust will yield. We
find a relationship between the depth of a crustal fracture, the
breaking strain and the corresponding energy release, and deduce a
characteristic field strength associated with crustquakes. We argue
that magnetically-induced crustquakes could power even the most luminous magnetar
phenomena, contrary to previous suggestions, as well as operate in NSs with
less exceptional inferred dipole magnetic field strengths.

\section{Magnetic-field equilibrium sequences}
\label{equilibria}

Around a day into its life a neutron star begins to form a crust,
crystallising gradually from the inside out over the course of the
following century\footnote{\red{See, e.g., \citet{ruderman_1968} for an
  early discussion of this; \citet{gnedin} and references therein for
  the theory of crustal thermal relaxation; and \citet{krueger} for a
  figure of how different regions freeze into a crust over time.}}. Before the crust has even
begun to form, however, it is reasonable to expect the magnetic field to have
reached an equilibrium with the fluid star, since the timescale of
this process will be the same order of magnitude as an Alfv\'en-wave
crossing time (around a second for typical NS parameters and a
$10^{14}$ G field; shorter for stronger fields). The crust will thus
freeze in a relaxed state threaded by its early-stage magnetic field;
in the absence of shear stresses the equilibrium description of this
phase will just be that of a magnetised fluid body (left panel of
figure \ref{crust-crack}). Over time the star will gradually lose magnetic
energy through secular decay processes; see section \ref{decay}. The magnetic field will want to
adjust to a new fluid equilibrium, but its rearrangement will be
inhibited by the crust's rigidity (middle plot of figure
\ref{crust-crack}). The magnetically-induced stresses in the crust
will thus grow over time, and eventually exceed the elastic yield
value; when this happens the crust will break in the region where its
breaking strain has been exceeded, and the field will be able to
return (locally) to its fluid equilibrium configuration, depicted in
the right-hand panel of figure \ref{crust-crack}. The stress that
builds up in a NS's crust will thus be 
sourced by the difference between the field configuration present when
the crust froze, and the magnetic field's desired present equilibrium,
which it is prevented from reaching by shear stresses. Both the `before'
and the (desired) `after' magnetic-field configurations are therefore fluid
equilibria, so by comparing two such equilibria with different values of
magnetic energy we can determine the expected stresses built up in an
elastic crust. A quantitative description of the above scenario
is given in section \ref{strain-deriv}, and its potential shortcomings
are discussed in the following subsection, \ref{caveats}.

To explore the possible range of these `before' and `after' magnetic-field
configurations during the evolution of a highly-magnetised neutron
star, we consider three classes of neutron-star model:
accounting for the possibilities that the core protons are
superconducting or not, and considering a scenario where the star has
a magnetar-like magnetosphere in equilibrium with its
interior field (and matches smoothly to it at the stellar
surface). From these various plausible models of a NS's field
configuration we hope to look for universal features and also
possible differences in how the crust breaks which could be used to
distinguish between them.

Our model NS is composed of protons, neutrons and electrons, but
the electrons have negligible inertia and their chemical potential can
simply be added as an extra contribution to that of the protons. We
are then left with a two-fluid core of protons and superfluid neutrons, matched at $0.9$ times the stellar
radius $R_*$ to a non-superconducting and unstrained crust. We capture
these features by confining the neutron fluid to the region between the
centre and $0.9R_*$, while having the proton fluid extend from the centre
out to the stellar surface, so that the shell from $0.9R_*$ to $R_*$
is a single-fluid region. Since a relaxed elastic medium obeys the
same equilibrium equations as a fluid, we can thus regard the proton
fluid in this outer single-fluid region as a `crust'
\citep{prix_novcom}. The equation of state we choose is effectively a double
polytrope \citep{LAG}, \red{setting the proton and neutron polytropic
  indices $N_\textrm{p},N_\textrm{n}$ to values of $1.5$ and $1.0$
  respectively, to mimic a `realistic' core proton-fraction profile in the
  core (e.g. that of \citet{douchin})}. Since the polytropic indices of the two
fluids are different, the stellar models have composition-gradient
stratification. The neutron-density profile has, however,
negligible impact on these configurations. \red{Finally, although it would naturally be
  more desirable to work directly with a tabulated equation of state,
  instead of our double-polytrope approximation to one, we do not
  believe that doing so would have any serious impact on our
  results; see the discussion in section \ref{crust_props}.}

The code we use to calculate equilibria works in dimensionless units, and physical values
given here come from redimensionalising code results to one
particular model star of $1.4$ solar masses and with fixed neutron and proton
polytropic constants of
$k_\textrm{n}=5.65\times 10^4\ \textrm{g}^{-1}\textrm{cm}^{5}\textrm{s}^{-2}$
and
$k_\textrm{p}=2.74\times 10^{10}\ \textrm{g}^{-2/3}\textrm{cm}^{4}\textrm{s}^{-2}$
respectively. For our chosen neutron and proton density profiles these values produce a star with
radius of 12 km (varying very slightly with field
strength). Note that since all our models have the same mass and the same
equation of state (i.e. fixed polytropic indices and constants), they
correspond to the same physical star --
allowing for a direct comparison between different models.

In all cases we are interested in mixed poloidal-toroidal
magnetic-field configurations, since these are the most generic
models and also the most likely to be stable
\citep{tayler_mix}. Although the toroidal-field component can be
locally strong in our
models, its contribution to the total magnetic energy is always small compared
with the poloidal one. The equilibrium models we consider here are all
chosen to have the strongest possible toroidal component; as we will
see later in section 4, this allows us to put an upper limit on how
readily the crust will yield.

The key differences between our three classes of model come from the
form of the magnetic force $\fmag$, and the electric current distribution; we
discuss each case next and show example field configurations, all with
a polar-cap field strength $B_p=6.0\times 10^{14}$ G for direct comparison.

\subsection{Normal core protons, vacuum exterior}

\begin{figure}
\begin{center}
\begin{minipage}[c]{0.8\linewidth}
\includegraphics[width=\linewidth]{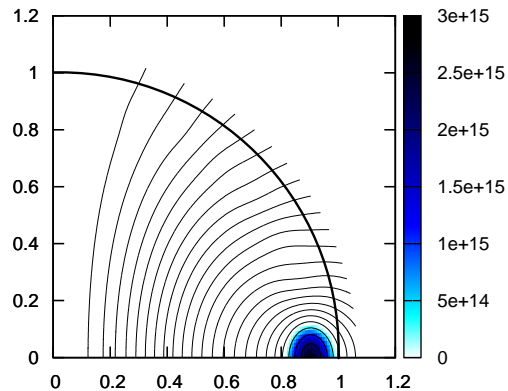}
\end{minipage}
\caption{\label{norm_6e14}
               Magnetic-field configuration for a model magnetar with
               a polar-cap field strength $B_p=6.0\times 10^{14}$
               G and a total magnetic energy of $6.8\times 10^{47}$ erg. The core is a two-fluid system of superfluid
               neutrons and normal protons, matched to a normal crust
               at a dimensionless radius $r/R_*=0.9$. The thick black
               arc at $r/R_*=1$ represents the stellar surface. We plot
               the poloidal-field lines, denoting the direction of this field component, whilst
               the colour scale shows the magnitude of the toroidal
               component, whose direction is azimuthal -- into/out of
               the page.}
\end{center}
\end{figure}

This is the simplest case, where both the core protons and the crust are subject to the familiar
Lorentz force for normal (non-superconducting) matter:
\beq \label{lorentz}
\fmag = \frac{1}{4\pi}(\curl\bB)\times\bB,
\eeq
where $\bB$ is the magnetic field.
Note that the neutron fluid does not feel any magnetic force. We
assume that the exterior of the star is a vacuum, with no
charged particles able to carry an electric current, so that
Amp\`ere's law simply imposes a restriction on the form of the
external magnetic field $\bB_\textrm{ext}$:
\beq
\curl\bB_{\textrm{ext}} = 0.
\eeq
An alternative way to look at this condition is that there
\emph{could} be magnetospheric currents, but that they do not
communicate with the interior and therefore do not affect its
equilibrium\footnote{The converse assumption -- that the interior
  field does not influence the exterior -- is standard in pulsar
  magnetosphere modelling; see the discussion in \citet{GLA}.}.
One could justify this rather simplistic model by suggesting that the magnetic field in
a magnetar's core is strong enough to break proton superconductivity
\citep{baym_pp,sin_sed}, so that the normal-matter
equations would apply. One key motivation for us, however, is that
it allows us to produce configurations with stronger toroidal
components than in our other cases;
see figure \ref{norm_6e14}.  We believe the reason for this
to be numerical rather than physical -- our code's iterative scheme converges to
strong-toroidal-field solutions more readily in this case than for the
other two models considered in this paper. This class of model is constructed using the
techniques described in \citet{LAG}, although the resultant field
configurations are not dissimilar to those of single-fluid models.

\subsection{Superconducting core protons, vacuum exterior}

\begin{figure}
\begin{center}
\begin{minipage}[c]{0.8\linewidth}
\includegraphics[width=\linewidth]{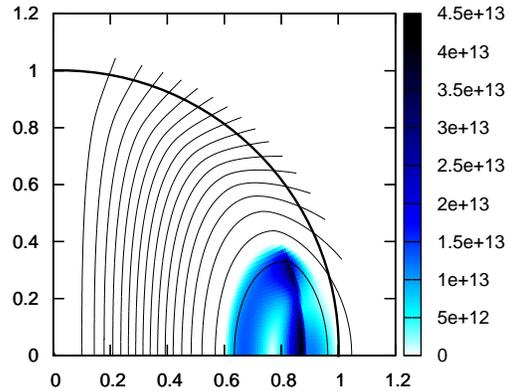}
\end{minipage}
\caption{\label{sc_6e14}
               Magnetic-field configuration for a model NS with
               $B_p=6.0\times 10^{14}$ G, with a
               superfluid-superconducting core matched to a normal
               crust. The total magnetic energy for this model is
               $3.1\times 10^{48}$ erg. The crust-core boundary and surface are at
               dimensionless radii of $0.9$ and $1.0$ as before; and
               again, we plot poloidal field lines in black and
               toroidal-field magnitude with the colour scale. Note
               the weakness of the toroidal component compared with
               the normal-matter model in figure \ref{norm_6e14}.}
\end{center}
\end{figure}

Our next class of equilibrium models are constructed as described in
\citet{sc_eqm_paper}. These consist of a core of superfluid neutrons
and type-II superconducting protons, matched to a normal crust. 
In the crust, the magnetic field is smoothly distributed (on a
microscopic scale) and the
magnetic force is just the Lorentz force \eqref{lorentz}, acting on the entire crust. In the core, by contrast, the
effect of proton superconductivity is to quantise the field into
an array of thin fluxtubes; on the macroscopic level, this produces a
different magnetic force. Unlike the Lorentz force, which depends only
on the macroscopic field $\bB$, the magnetic force for a type-II
superconductor also involves the \emph{lower critical field} ${\bf
  H}_{c1}$, related to the magnetic field along fluxtubes \citep{easson_peth,akgun_wass,GAS}. This latter field is
parallel to $\bB$ and proportional to the local proton density: in the
centre, where the proton density is highest, it reaches $10^{16}$ G,
but is on average around $10^{15}$ G within the core, irrespective of the value of
$\bB$. The most important feature governing these equilibria is the difference in
the form of the magnetic force for the core and crust:
\beq \label{mag_forces}
\fmag = \left\{
    \begin{split}
      & \frac{1}{4\pi}(\curl{\bf H}_{c1})\times\bB-\frac{\rhop}{4\pi}\nabla\brac{B\pd{H_{c1}}{\rhop}}
                       & \ \ \textrm{(core)} \\
      & \frac{1}{4\pi}(\curl\bB)\times\bB
                       & \ \ \textrm{(crust)} \\  
    \end{split}
    \right.
\eeq
Our models assume the core and crustal fields match without any
current sheet in this region, and as for the models described in the
previous subsection do not have any exterior current. An example of a
model with core proton superconductivity is shown in figure \ref{sc_6e14}.

\subsection{Normal core protons, magnetosphere}

\begin{figure}
\begin{center}
\begin{minipage}[c]{0.8\linewidth}
\psfrag{before}{initial state}
\psfrag{after}{after magnetospheric decay}
\includegraphics[width=\linewidth]{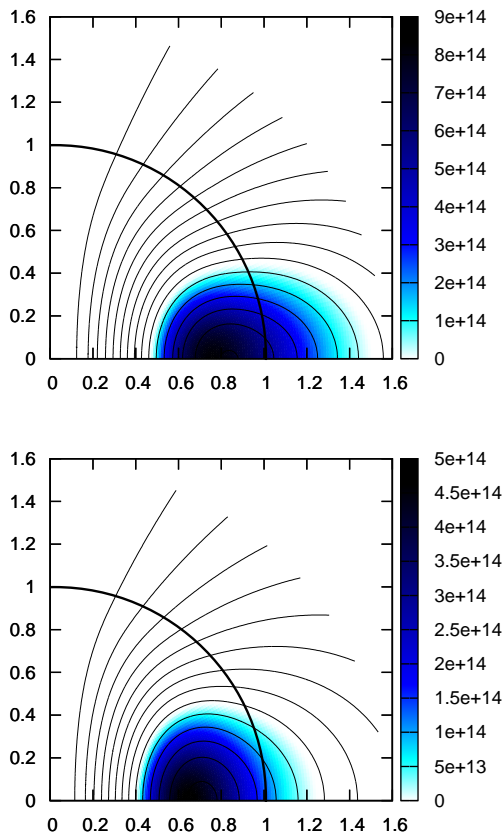}
\end{minipage}
\caption{\label{magneto_6e14}
               Two magnetic-field configurations for a normal-matter
               NS with a current-carrying corona and
               $B_p=6.0\times 10^{14}$ G, but with the lower plot having
               $2.5\times 10^{46}$ erg less magnetic energy. All of
               this energy has been taken out of the toroidal
               component, visibly altering the magnetosphere. The
               toroidal component attains a maximum value greater than
               that of the model in figure \ref{sc_6e14}, but still less
               than that in figure \ref{norm_6e14}. The lower model
               has $5.4\times 10^{47}$ erg of magnetic energy.}
\end{center}
\end{figure}

These equilibria are constructed in the same way as the
models with a normal core, except that we now allow for a
toroidal-field component that extends outside the star. This is
sourced by a poloidal electric current in a magnetosphere of charged
particles, located in a lobe around the equator as argued for by
\citet{belo_thomp}. Outside the lobe region there is vacuum,
where the field obeys $\curl\bB_{\textrm{ext}}=0$, but within it there
is a force-free region with
\beq
\curl\bB_\textrm{ext}=\alpha\bB_\textrm{ext},
\eeq
where $\alpha$ is a function constant along magnetic-field lines,
governing the distribution of magnetospheric current; see
\citet{GLA} for details on the method of solution for these
configurations. In figure \ref{magneto_6e14} we plot two such models
of NSs in dynamical equilibrium with their magnetosphere, both
with $B_p=6.0\times 10^{14}$ G, but with $2.5\times 10^{46}$ erg of magnetic energy
removed from the toroidal component in the second plot, illustrating
how the magnetosphere rearranges in this case.

At this point it is worth speculating about scenarios for the
formation of an equatorial corona of current-carrying plasma, although
this is not the focus of our work. The standard
argument for the formation of such a corona \citep{belo_thomp} assumes
the interior field evolves in such a way that it wishes to `eject
magnetic helicity' -- equivalently, to induce an electric current in
the environs of the star. For a mature NS this process cannot
happen immediately, but initially results in crustal stresses building
-- when these are released the initially poloidal field is twisted in an azimuthal direction,
thus generating a toroidal component.

As shown in \citet{GLA}, given a
sufficiently dense corona of charged particles, the star can form a
magnetosphere which is in dynamical equilibrium with the internal
field and hence supported by a relaxed crust, as opposed to one which
has to shear to generate the field. If we assume the toroidal
component is always confined to the same flux surface (poloidal field
line), then the decay of this field would cause the magnetosphere to change
shape, moving in towards the crust. As before, the magnetic flux's
inward motion would initially be inhibited by shear forces, but at a
later stage the induced stresses could grow large enough to break the crust. Figure
\ref{magneto_6e14} assumes a scenario like this, where the
configuration of the upper panel decays into that of the lower panel.

One could also view the panels in reverse, however, where an internal toroidal
field wishes to rise out of the star for whatever reason, but is again
inhibited by the crust. Clearly one cannot view the specific
configurations of figure \ref{magneto_6e14} as representing this scenario,
since that would require an increase in magnetic energy, but
qualitatively similar solutions with decreasing magnetic energy could
be constructed. In terms of a
changing global equilibrium this way round seems less likely, but
it does broadly represent the corona-formation mechanism discussed in
\citet{belo_thomp} and \citet{belo09}. Note that the strain patterns that would
build by running the scenario in this order would be the same as in
the reverse order, however, since the strain/yield criterion of
section \ref{strain-deriv} remains the same if the `before' and
`after' configurations are swapped around.

\section{Field decay}
\label{decay}

The fact that NS magnetic fields \emph{do} decay is well-established, from both
theoretical and observational study; our knowledge of the relative importance of
different decay mechanisms, and their corresponding timescales, is
nevertheless surprisingly incomplete. If the activity of young NSs like magnetars is
powered by field decay, however, there must be at least one
rather rapidly-acting decay mechanism. We therefore consider it
reasonable to \emph{assume} magnetically-induced stresses will
build in a NS crust on a timescale short enough to be astrophysically
relevant, even if current theoretical uncertainties
prevent us from pinpointing the mechanism(s) which will most readily
build these stresses. Here we briefly
review the literature on magnetic-field evolution to highlight the most promising
mechanisms for relatively rapid changes in a NS's field.

The most familiar source of magnetic field dissipation is Ohmic decay, which in
terrestrial materials and the neutron-star crust is the macroscopic
result of electrons scattering off a solid material's ion lattice, thus
heating it and reducing the electric current. Ohmic decay operates
more rapidly on small-scale fields than large-scale ones. In the crust
of a NS the separate process of Hall
drift acts to redistribute the magnetic flux into structures of
progressively shorter lengthscales \citep{gold_reis}; although this
process is not itself dissipative it aids Ohmic decay,
which acts more rapidly on small-scale magnetic fields.

For the core, many studies have argued that
the evolution is likely to be very slow. Ohmic decay itself must be
restricted to the thin cores of normal protons at the centre of
fluxtubes -- these cores comprise a minute volume of the NS core,
which is otherwise in a superconducting state, and so this decay
mechanism is expected to be extremely slow \citep{baym_pp2}. Ambipolar diffusion --
a drift of the charged particles, and hence the magnetic field, with
respect to the neutrons -- is both dissipative and acts to move the
core magnetic field outwards into the crust \citep{gold_reis}. However,
accounting for the superfluid state of the neutrons drastically
increases its timescale \citep{GJS}.

For magnetic
fields $B<H_{c1}\approx 10^{15}$ G, the Meissner effect is expected to
expel core magnetic flux to the crust-core boundary, a region of
(probably) higher electrical resistivity and hence faster Ohmic decay. More precisely, the Meissner effect
dictates that the eventual equilibrium state of the field will be one where it
is exponentially screened from the core over some short lengthscale --
it does not specify the dynamical mechanism which might achieve
this, nor the timescale. Different mechanisms have been invoked for
the transport of magnetic flux out of the core. The fluxtubes may move
out of the core through mutual self-repulsion \citep{kocharovsky},
driven by a buoyancy force \citep{mus_tsyg,wendell,harrison91,jones91}, or dragged by the 
outwardly-moving neutron vortices as the star's rotation rate
decreases \citep{ding_cc}. The result of these various studies is an assortment of prospective timescales
for field decay which range over at least eight (!) orders of magnitude
($10^4-10^{12}$ yr). Nonetheless the consensus, inasmuch as there is
one, points to a rather slow core evolution and suggests that observed
field decay is crustal in origin.

Slow core-field evolution may, however, be contradicted by the
observation that young NSs like magnetars are able to build and
release huge stresses: the most energetic giant flare ($\sim\!
10^{46}\ \textrm{erg}$) came from a magnetar believed to be under a
thousand years old \citep{palmer,tendulkar}. Alternatively, instead of
being the result of secular stress build-up, magnetar giant flares may be
the manifestation of a rapidly-acting hydromagnetic instability \citep{thom_dunc96,ioka}
-- although that in itself requires the instability to be somehow
suppressed until some critical point, and therefore one might again have
to invoke the build-up of crustal stresses. There is clearly more
work to be done in attempting to achieve some kind of consensus on the
role of magnetic field decay in NS phenomena -- but if crustquakes induce
magnetar activity, as discussed in this paper, we may in fact be able to use
\emph{observations} to determine a core field decay timescale and hence
reduce the discordance of the theoretical models.

\section{Magnetically-induced crustquakes}

\subsection{Crustal properties}
\label{crust_props}

\red{To obtain quantitative results about how a magnetic field can act
  to strain and eventually break a neutron-star crust, we need a
  realistic model of this region -- in particular, for the crustal shear
  modulus and breaking strain. In our equilibrium models, described in
  section \ref{equilibria}, we used a
  double-polytrope equation of state designed to mimic a `realistic'
  core proton fraction, but unfortunately this results in an unrealistically low density crust.} The crustal
density distribution does not have a strong impact on the magnetic
field configuration, but is important for calculating a reasonable
shear-modulus profile. Accordingly, we choose to take quantities from
a tabulated, `realistic', equation of state \citep{douchin}, and by doing so we
can take advantage of a recent shear-modulus fitting formula based on
the results of molecular-dynamics simulations \citep{horo_hugh}.

Using a polytropic crust model to calculate magnetic-field equilibria, but then adopting
parameters from a tabulated equation of state to calculate the shear
modulus, is clearly not consistent. Nonetheless, we
argue next that the level of inconsistency is justifiable to our order
of working. We use our equilibrium calculations solely to get models
of the magnetic field and not, for example, pressure or density
profiles. Had we employed the Douchin-Haensel equation of state
consistently throughout this work -- i.e. for our equilibrium
calculations too -- we would have obtained somewhat different
magnetic-field distributions. The degree of inconsistency in our
approach in this paper, therefore, depends on the difference between
equilibria calculated using our polytropic models and those calculated
with the equation of state of \citet{douchin} -- and this difference
should be small, since the dependence of magnetic-field distributions
on the stellar equation of state is in fact quite weak \citep{yosh_yosh_eri}.

\begin{figure}
\begin{center}
\begin{minipage}[c]{\linewidth}
\psfrag{log_mu}{$\displaystyle\log\brac{\frac{\mu}{\textrm{dyn\ cm}^{-2}}}$}
\psfrag{r_dimless}{$r/R_*$}
\includegraphics[width=\linewidth]{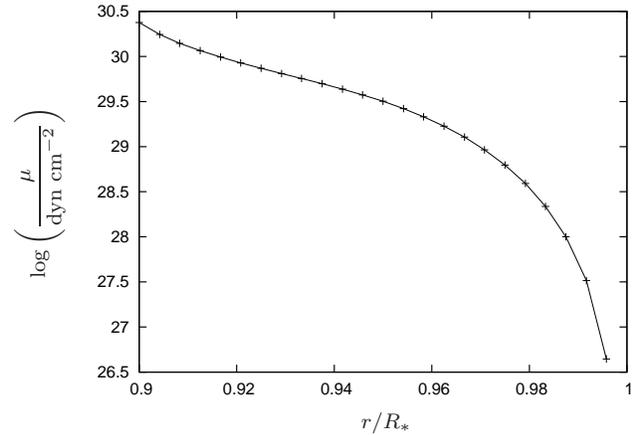}
\end{minipage}
\caption{\label{shear_mod}
               The profile of the shear modulus $\mu$ throughout
               the crust, calculated using equation
               \eqref{HH_shear-mod} -- a fitting formula based on the
               results of molecular-dynamics simulations \citep{horo_hugh}. The required crustal input
               quantities (for example, the variation of atomic number within
               the crust) come from polynomial fits to the
               tabulated equation of state of \citet{douchin}, and
               our magnetar temperature profile is taken from
               \citet{kaminker_2009}.}
\end{center}
\end{figure}

\subsubsection{\red{Shear modulus}}

From the Douchin-Haensel tabulated equation of state we make simple
polynomial fits to the radial dependence of the baryon
number $n_b$, atomic weight $A$, atomic number $Z$ and free neutron
fraction $x_\textrm{n}^\textrm{free}$ in the crust. We fit our temperature profile to
results for a 1000-year-old magnetar from \citet{kaminker_2009} (see
their figure 6; we use their profile for the lower of the two heat
intensities, with a heat source at the top of the inner crust). The
maximum temperature slightly exceeds $10^9$ K. Our
fitting formulae approximate crustal parameters over the density range
$0.05\rho_\textrm{cc}<\rho<\rho_\textrm{cc}$ (where $\rho_\textrm{cc}$ is the
density at the crust-core boundary) and may deviate from the correct
behaviour below this density. On our numerical grid the crust is
covered by $24$ radial points, meaning that our fitting formulae are
designed to approximate all but the outermost four points -- precise
enough for our purposes.

To calculate crustal properties, we first note that the ion number
density in the crust $n_i=n_b(1-x_\textrm{n}^\textrm{free})/A$, from which we define
the ion sphere radius $a_i=(4\pi n_i/3)^{-1/3}$. The Coulomb coupling
parameter $\Gamma$ is then given by
\beq
\Gamma = \frac{(Ze)^2}{a_i T},
\eeq
where $e$ is the elementary charge (i.e. of a proton).
From the various crustal properties discussed above, we are now in a position to determine
the shear modulus $\mu$ of our model NS crust
using the formula from \citet{horo_hugh}:
\beq \label{HH_shear-mod}
\mu = \brac{0.1106-\frac{28.7}{\Gamma^{1.3}}} \frac{n_i}{a_i}(Ze)^2.
\eeq
The resulting shear modulus profile we use is shown in figure
\ref{shear_mod}. At the innermost crustal gridpoint $\mu=2.4\times
10^{30}$ dyn cm${}^{-2}$ -- a little higher than the crust-core value 
of $\mu=1.8\times 10^{30}$ dyn cm${}^{-2}$ from \citet{hoffman_heyl} and the classic
estimate of $\mu\approx 10^{30}$ dyn cm${}^{-2}$ \citep{ruderman_1969}.

\subsubsection{\red{Breaking strain}}
\label{breaking_strain}

\red{Recent molecular-dynamics simulations indicate that the neutron-star
crust is considerably stronger than previously thought
\citep{horo_kad,hoffman_heyl}, with a
breaking strain $\sigma_\textrm{max}$ around $0.1$ (dimensionless,
since a strain is a fractional deformation; a ratio of two
lengths). $\sigma_\textrm{max}$ is essentially temperature-independent as long
as one is well above the melting temperature (whose value corresponds to
$\Gamma\approx 175$); it is also independent of density, except perhaps
in a narrow region of `nuclear pasta' at the crust-core boundary
\citep{ravenhall}; and neither impurities nor strain rate have a
significant impact on it. Accordingly, taking the
breaking strain as constant is a good first approximation
\citep{horo_private}. Note that the breaking \emph{stress}, by
contrast, is a dimensional quantity (with units of pressure) and has significant
variation within the crust \citep{chug_horo}. In this paper we adopt two
canonical values for the breaking strain: $\sigma_\textrm{max}=0.1$ to
reflect recent simulations, and $\sigma_\textrm{max}=0.001$ to compare
with earlier work.}

\subsection{A criterion for magnetically-induced crustquakes}
\label{strain-deriv}

We are interested in how magnetic field decay/rearrangement causes
strain to build in a neutron star's crust, and where and when this
strain might finally cause the crust to break. Since there is no 
reason to expect the magnetic field to be uniform -- or to
decay/rearrange uniformly -- the built-up strains will vary
greatly within the crust. Previous crust-breaking criteria based on global
estimates \citep{thom_dunc95,hoffman_heyl} are therefore not only
somewhat crude, but also give no idea of \emph{where} the crust will
fail. We aim to improve on these by using a criterion, which we derive
next, accounting for the \emph{local} changes in magnitude and direction of the
field.

To simplify the algebra in the derivation which follows, we use
standard tensor index notation, denoting tensor indices with $i$ and
$j$. We start with the general stress tensor for the crust in our
model:
\beq
\tau_{ij} = -pg_{ij} + \mu\sigma_{ij} + \clM_{ij},
\eeq
where $p$ is fluid pressure, $g_{ij}$ the flat-space 3-metric,
$\sigma_{ij}$ the elastic strain tensor and $\clM_{ij}$ the Maxwell
(magnetic) stress tensor. In this problem we are only considering equilibrium
configurations --- either strained or unstrained --- so the sum of the
stresses should balance: $\tau_{ij}=0_{ij}$, where $0_{ij}=0\ \forall\{i,j\}$.

We assume the NS's crust freezes in a relaxed state, with
a certain magnetic energy and polar-cap field strength; quantities
pertaining to this state will be denoted with a subscript or
superscript zero in the following derivation. With no shear
forces present, the equilibrium at this stage is
that of a fluid body:
\beq
0_{ij} = -p^0 g_{ij} + \clM^0_{ij}.
\eeq
Over the star's lifetime, different secular processes (see previous
section) act to reduce the magnetic energy, so that the star's evolution can
be described by a sequence of quasi-static equilibria, with
incrementally smaller values of magnetic energy. These are no longer fluid equilibria,
however, as the crust resists any adjustment of the magnetic field by
balancing the Lorentz forces by its elastic shear force:
\beq
0_{ij} = -pg_{ij} + \mu\sigma_{ij} + \clM_{ij}.
\eeq
The magnetically-induced change to the fluid pressure $p$ will be
tiny, and so the difference between its initial value and that at a
later time may safely be neglected, i.e. $p^0-p\approx 0$. The strain
in the crust is thus entirely sourced by the difference in the Maxwell
stress tensor between initial and later field configurations:
\beq \label{maxwell-diff}
\mu\sigma_{ij} = \clM^0_{ij} - \clM_{ij}.
\eeq
The magnetically-induced stresses in the crust gradually grow, and are
largest where the field wishes to adjust the most. For sufficiently strong
magnetic fields and sufficient readjustment, the crust will yield in
some region, allowing the magnetic field in
the affected region to return to a fluid equilibrium; recall the
cartoon in figure \ref{crust-crack}.

To proceed we need the explicit form of $\clM_{ij}$. Since the crust
is not superconducting this is the familiar Maxwell stress tensor:
\beq \label{maxwell-tensor}
\clM_{ij} = \frac{1}{4\pi}\brac{B_i B_j - \frac{1}{2}B^2\delta_{ij} }.
\eeq
Note that taking the divergence of this tensor gives:
\beq
\div\clM = \frac{(\bB\cdot\nabla)\bB}{4\pi} - \frac{\nabla B^2}{8\pi},
\eeq
the Lorentz force, as expected.

The von Mises criterion predicts that an isotropic elastic medium will
yield when
\beq \label{von_mises_orig}
\sqrt{ \textstyle{\frac{1}{2}}\sigma_{ij}\sigma^{ij} } \geq \sigma_\textrm{max}.
\eeq
This is not, strictly
speaking, a criterion for breaking; `yield' means only that the crust
ceases to respond elastically to additional strains, but may enter a regime of plastic
flow before actually breaking. We ignore the distinction between these
two responses for now, and use the terms `yield' and `break'
interchangeably. For the purposes of our work the distinction is not so
important, as we anticipate both breaking and plastic flow to release
the same total amount of pent-up magnetic energy, but perhaps in
characteristically different ways and over different timescales; see
section \ref{discussion}.

Now, from equations \eqref{maxwell-diff} and \eqref{maxwell-tensor}:
\beqa
\sigma_{ij}\sigma^{ij} 
 &= \frac{1}{\mu^2}
        \brac{  \clM^0_{ij}\clM^{ij}_0 + \clM_{ij}\clM^{ij}
                    - \clM^0_{ij}\clM^{ij} - \clM_{ij}\clM^{ij}_0  } \nn \\
 &= \frac{1}{64\pi^2\mu^2}
        \brac{ 2B^2 B_0^2 + 3B^4 + 3B_0^4 - 8(\bB\cdot\bB_0)^2 }.
\end{align}
The von Mises criterion \eqref{von_mises_orig} applied to the case of
crust-yielding sourced by a changing magnetic-field equilibrium is therefore:
\beq
\frac{1}{8\pi\mu}\sqrt{ B^2 B_0^2 + \textstyle{\frac{3}{2}}B^4 + \textstyle{\frac{3}{2}}B_0^4
                                     - 4(\bB\cdot\bB_0)^2 } \geq \sigma_\textrm{max}.
\eeq
Since we will explore varying the breaking strain, we will use the
following quantity in strain plots:
\beq \label{von_mises_mag}
\frac{\sqrt{ B^2 B_0^2 + \textstyle{\frac{3}{2}}B^4 
          + \textstyle{\frac{3}{2}}B_0^4 - 4(\bB\cdot\bB_0)^2 }}{8\pi\mu\sigma_\textrm{max}}.
\eeq
Accordingly, we expect any regions in the crust where this quantity
exceeds unity to break. We consider the validity of our crustquake model
and alternatives to it in the next subsection, and then present our results.

\subsection{Validity of our crustquake criterion}
\label{caveats}

In this paper we aim to take a commonly-invoked idea of magnetic field decay
driving crustquakes and put it on a firm quantitative footing. Our
approach, in summary, is to study how a changing magnetic-field
equilibrium strains a NS's crust. We do not perform time-dependent
simulations of this process, so we cannot actually simulate a fracture
event -- instead, we use the von Mises yield criterion to check which
regions of the crust have exceeded the breaking strain, and infer that
those regions will yield. We have in
mind a scenario where a substantial region of the crust fails
collectively in a fracture -- which appears contradictory
to a recent suggestion that crack propagation, and hence mechanical
failure, is inhibited in magnetised NS crusts by the Lorentz force
\citep{lev_lyu}. We are not considering mechanical failures with
arbitrary geometry, however, but ones which are \emph{induced} by the
Lorentz force and therefore are dictated by the magnetic-field
geometry rather than impeded by it. Nonetheless, even if the crust
fails gradually in small regions and/or enters a regime of plastic flow \citep{jones03,belo_lev}, the results we
present should still represent the total energy output over the
yield process.

We assume shear stresses are sourced solely by the crust resisting the
rearrangement of the star's hydromagnetic equilibrium. This is in the
same spirit as \citet{braith_spruit}, although their approach was to isolate one
piece of the Lorentz force to diagnose the build-up of stress, whereas
we have derived a tensor-based yield criterion which follows
rigorously from elasticity theory. By comparing hydromagnetic equilibria, we
are neglecting the separate secular evolution of the star's field, and
in particular that of the crust \citep{pons_mu}; the only
role of any dissipative effect in our models is to induce
the field to rearrange into a new equilibrium. Our study is therefore
complementary to the crustquake modelling of \citet{perna_pons}, who
\emph{did} look at the build-up of crustal stresses due to magneto-thermal evolution in
the crust, but neglected any effects related to changes in
the star's global equilibrium.

\begin{figure*}
\begin{center}
\begin{minipage}[c]{\linewidth}
\psfrag{supercon}{core superconductivity}
\psfrag{normal}{normal core}
\psfrag{vacuum}{vacuum exterior}
\psfrag{magnetosphere}{magnetosphere}
\psfrag{high-sigma}{high-$\sigma_{\textrm{max}}$}
\psfrag{low-sigma}{low-$\sigma_{\textrm{max}}$}
\includegraphics[width=\linewidth]{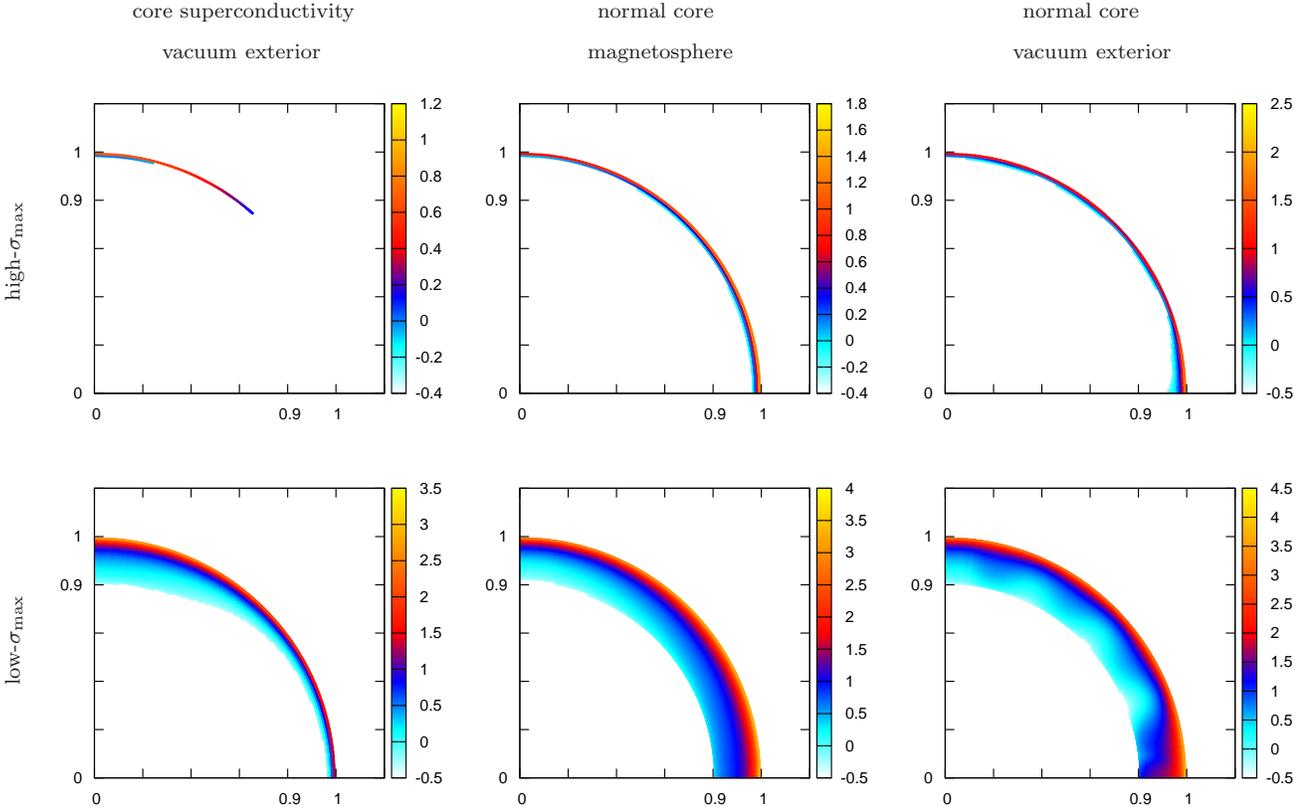}
\end{minipage}
\caption{\label{strains}
               Logarithmic plots of the ratio of magnetic strain to
               breaking strain within a neutron-star crust;
               when the ratio exceeds unity (i.e. zero for this
               logarithmic plot) the crust should break. The colour
               scale shows regions where the ratio is $0.5$ or
               greater, corresponding to $-0.3$ or greater on the
               logarithmic scale, reflecting the fact that a real
               NS crust's crystalline lattice may contain flaws and
               impurities which cause it to break before reaching the
               limit for a pure crust. We plot the crust at twice its
               actual thickness to show strain patterns more clearly. All
               plots show strain built up in NSs with a
               present-day field strength $B_p=6.0\times 10^{14}$ G,
               after the loss of $2.5\times 10^{46}$ erg of magnetic
               energy. This loss represents $0.80\%$ of the
               present-day total magnetic energy for the left-hand
               plots (superconducting core protons and a vacuum
               exterior), $4.7\%$ for the middle plots (normal core
               protons, non-vacuum exterior) and $3.6\%$ for the right-hand
               plots (normal core protons, vacuum exterior). The top row shows results for a very strong crust, with
               a breaking strain $\sigma_\textrm{max}=0.1$; the bottom row is the same
               set of configurations but assuming a more
               `traditional' value of $\sigma_\textrm{max}=0.001$. Our
               models show that a NS crust yields most easily if the
               star has a locally strong toroidal-field component,
               with the failure occurring in the outer equatorial
               region first.}
\end{center}
\end{figure*}

In addition to the potential role played by the secular
field-rearrangement processes in the crust, one other potential
concern is the inherent degeneracy in picking sequences of equilibria to
represent snapshots of the rearrangement of a NS's decaying magnetic
field. Since this process is dissipative, there is no obvious quantity
to hold constant -- in contrast with, for example, the case of
accretion-driven burial of a NS's magnetic field \citep{payne_mel}.
Although we rescale our numerical results to one specific physical NS
($1.4$ solar masses and a radius of $12$ 
km), our picture of a sequence of equilibria as snapshots of a secular
evolution is therefore not self-consistent. Somewhat arbitrarily, we assume that the ratio of poloidal
to toroidal components remains constant for our models with only
interior currents, whilst assuming that in our `magnetosphere' models
the exterior current decays most quickly, thus predominantly reducing
the toroidal component (which is partially sourced by these exterior
currents). Ideally one would verify these assumptions with a full
magneto-thermal evolution of the coupled core-crust-magnetosphere
system, but the technology to perform such simulations is not yet
available. For now, we believe the work presented in this paper to
be as complete as is currently possible.

\subsection{Strain patterns in a neutron-star crust}

In figure \ref{strains} we plot the strain patterns that would develop in a
NS crust after a period in which $2.5\times 10^{46}$ erg of
magnetic energy has decayed, assuming the crust's initial state was
relaxed. In all cases the final, `present-day' polar cap field strength is taken to
be $B_p=6.0\times 10^{14}$ G. For clarity the $0.1R_*$-thick crust
($1.2$ km for our models) has
been stretched linearly in the plot to appear at twice its actual thickness. We consider the three classes of model
described in section \ref{equilibria}: superconducting core protons
and a vacuum exterior; normal core protons and magnetospheric
currents; normal core protons and a vacuum exterior. We plot the
quantity from equation \eqref{von_mises_mag}, which is greater than
unity for regions of the crust which are expected to yield; since our
colourscale is logarithmic, zero represents the minimum value at which
the crust is expected to yield (given the caveats discussed in the
previous subsection). To include regions on the verge of breaking, we
also show parts of the crust where the quantity \eqref{von_mises_mag}
exceeds $0.5$. These parts may actually fail, rather than just being
on the verge of it, if the crustal lattice contains flaws/impurities
or if a large region fails collectively, for example. The colour scale shows how much
strain builds up in each region. The top row of plots assumes a very
strong crust, with $\sigma_\textrm{max}=0.1$, whilst the bottom row
uses $\sigma_\textrm{max}=0.001$ for comparison; see the discussion
in section \ref{breaking_strain}.

Superficially, figure \ref{strains} seems to suggest that
normal-matter models with a vacuum exterior are the most prone to
fracture, given a fixed loss of total magnetic energy. If we return to
the equilibrium models used to generate these plots (figures
\ref{norm_6e14}, \ref{sc_6e14} and \ref{magneto_6e14}), however, we see
that the comparison is not quite fair: the three classes of equilibria
have strikingly different toroidal-field strengths, with that of the
superconducting model being an order of magnitude weaker than the
other two. A more reliable conclusion to draw from our results is that a strong
toroidal-field component allows for the greatest build-up of stress
in a NS crust, in agreement with previous studies
\citep{thom_dunc95,pons_perna}. Our results are also distinct from this
earlier work, however, in that we anticipate the greatest stress
build-up -- and eventually a crustquake -- to occur in
a belt around the equator. By contrast, a poloidal-dominated field builds up
stresses more gradually, and in a region around the pole.

\subsection{Energy release and characteristic field strength for crustquakes}

One key question for any model of crustquakes is the amount of
energy that could be released in such an event. Here we compare
sequences of models to determine the relationship between the various
quantities in the problem: the energy release in a quake, the depth of
the `fracture' (i.e. the region which fails), the breaking strain and the polar-cap field
strength. For later comparison, we first quote the result of a
back-of-the-envelope calculation \citep{thom_dunc95} for crustquake
energy release:
\beq \label{TD95_estimate}
\frac{E_\textrm{out}}{10^{40}\textrm{ erg}}
 \sim 4\brac{\frac{l}{1\textrm{ km}}}^2
            \brac{\frac{B_\textrm{c}}{10^{15}\textrm{ G}}}^{-2}
            \brac{\frac{\sigma_\textrm{max}}{0.001}}^2,
\eeq
where $B_\textrm{c}$ is the crustal magnetic field.
Note that this estimate gives the energy released $E_\textrm{out}$ from the failure of
an \emph{area} of size $l^2$; we find the notion of energy release
from a volume more natural, since magnetic energy is a volume integral
over $B^2$.

For our results, we produce sequences of field configurations by fixing one equilibrium model, the
`after' (present-day) model with crustal strains sourced by the magnetic
field, and varying the other, `before' (original) configuration -- i.e. the
initial star with its relaxed crust. We assume the `before' field
has decayed into the `after' field -- so that the greater the
difference in magnetic energy between these models, the larger the
region of the crust that should be strained to the point of yielding. We also explore the effect of
varying the breaking strain and the `after' field strength. We
then compare the depth of the fracture in each case with
the magnetic-energy change in the region which fails, which we regard as the
energy released over the crustquake and denote $E_\textrm{quake}$.

As discussed in the previous subsection, our models with
normal core protons and vacuum exterior have the highest ratio of
toroidal-component maximum to polar-cap field strength. We believe
that equilibrium solutions with similarly high ratios do \emph{exist}
in other cases, in particular the case with core superconductivity,
but that our numerical scheme is simply less successful at converging
to them. In this section we will only consider the class of models
with a normal core and vacuum exterior, and will use the strongest
toroidal components we can, as before, since this seems to be
associated with the greatest build-up of strain. Given that we believe
similarly strong toroidal fields should exist in other cases, however,
the results presented here are intended to be representative of a favourable
crust-breaking scenario for \emph{any} model.

\begin{figure}
\begin{center}
\begin{minipage}[c]{\linewidth}
\psfrag{Emag}{$\displaystyle\frac{E_\textrm{quake}}{10^{45}\textrm{ erg}}$}
\psfrag{frac_depth}{$\displaystyle\frac{d}{R_\textrm{c}}$}
\includegraphics[width=\linewidth]{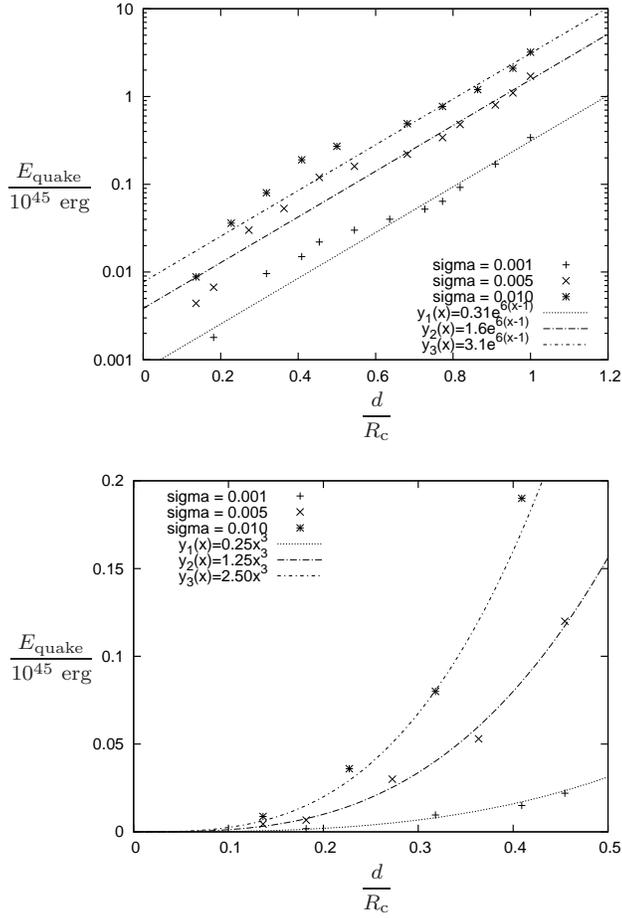}
\end{minipage}
\caption{\label{log-lin}
               The amount of energy released in a crustal
               fracture, as a function of fracture depth. Fixing
               the present-day polar-cap field strength as
               $B_p=3.0\times 10^{14}$ G we consider three different
               breaking strains, as labelled on the figure:
               $\sigma_\textrm{max}=0.001,0.005,0.01$. Top: for
               sufficiently deep fractures the relationship between
               depth and energy loss is approximately
               exponential (shown by the lines). Bottom: a zoomed-in
               version of the above shows that for more
               shallow fractures the relationship deviates
               from the exponential one and is better approximated by
               a cubic function.}
\end{center}
\end{figure}

We begin by fixing the present-day polar-cap field strength as
$3.0\times 10^{14}$ G and varying the initial field strength. We then
calculate the ratio of magnetically-induced strain $\sigma$ to
breaking strain $\sigma_\textrm{max}$ throughout the crust,
using equation \eqref{von_mises_mag}, to determine
what depth of region will fail according to the von Mises yield
criterion. The difference in magnetic energy between the
`before' and `after' equilibrium configurations, within the volume of
the crust which breaks, gives us the energy $E_\textrm{quake}$ released in such a
crustquake:
\beq \label{E_quake}
E_\textrm{quake} 
= \int\limits_{\sigma\geq\sigma_\textrm{max}}
          \frac{(B_0^2-B^2)}{8\pi}\ \textrm{d}V.
\eeq
Our results for the variation of energy release with fracture depth
are plotted in figure \ref{log-lin} for three different breaking
strains, \red{to allow us to check the dependence on this quantity too}. For fracture
depths exceeding around half the crustal thickness, we find that the data is fitted satisfactorily by
an exponential relation between energy release and depth; see top
panel. For more shallow fractures, however, a cubic fit is better
(bottom panel). Note that the exponential relation could not in any 
case be applicable at shallow depths, since it does not give the
correct limiting behaviour that if there is no fracture there can be
no energy release (i.e. the energy-versus-depth fit line must pass through the origin).

\begin{figure}
\begin{center}
\begin{minipage}[c]{\linewidth}
\psfrag{Emag}{$\displaystyle\frac{E_\textrm{quake}}{10^{45}\textrm{ erg}}$}
\psfrag{frac_depth}{$\displaystyle\frac{d}{R_\textrm{c}}$}
\includegraphics[width=\linewidth]{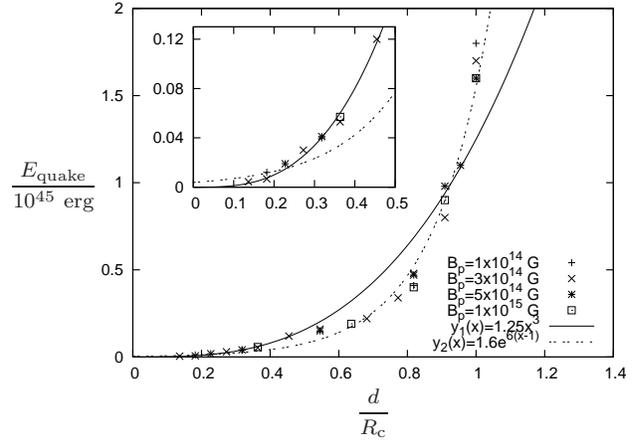}
\end{minipage}
\caption{\label{varyB}
               The relationship between fracture depth and energy
               release for four different present-day polar-cap field
               strengths, for a breaking strain of $0.005$. We see
               that the results appear to be completely independent of
               the field strength. The exponential fit is seen to
               approximate the behaviour for deep fractures and large
               energy release, whilst the cubic fit (inset) is more
               accurate for shallow fractures and smaller release of energy.}
\end{center}
\end{figure}

\red{Since equation \eqref{TD95_estimate} suggests our results may be
  dependent on the NS's field strength, we investigate this next}. In
figure \ref{varyB} we fix the breaking strain at $0.005$ and show
the variation of crustquake energy release with fracture depth for four different present-day polar-cap field strengths,
varying over an order of magnitude. The data points for the four
different field strengths all appear to lie along the same line, with
no evident variation with field strength. This is not actually so
surprising -- \red{whilst the stresses are \emph{induced} by the magnetic
  field, they are stored as elastic energy, so that crustquake energy
  release depends only on crustal properties: the volume of the crust which yields and
  the strain at which this occurs. The magnetic-field strength is
  likely to be important in affecting the rate of crustquake events, but such
  time-dependent behaviour is beyond the scope of this paper}.

As for figure
\ref{log-lin}, we see in figure \ref{varyB} that
the quake energy-depth relation appears to be exponential for deeper
fractures, and cubic for shallower fractures (see inset). Combining
our results from these last two figures and denoting the crustal
thickness by $R_\textrm{c}$, we find that the relation
between quake depth $d$ and energy release is \emph{independent} of the field
strength -- in contrast with earlier estimates -- and given by
\beq \label{E-d_deep}
\frac{E_\textrm{quake}}{10^{45}\textrm{ erg}}
 = 0.31\brac{\frac{\sigma_\textrm{max}}{0.001}}
            \exp\left[6(\textstyle\frac{d}{R_{\textrm{c}}}-1)\right]
\eeq
for deep fractures ($d\gtrsim 0.5R_\textrm{c}$), and
\beq \label{E-d_shallow}
\frac{E_{\textrm{quake}}}{10^{45}\textrm{ erg}}
  = 0.25\brac{\frac{\sigma_{\textrm{max}}}{0.001}}
             \brac{\frac{d}{R_\textrm{c}}}^3
\eeq
for more shallow ones.

Since we work in axisymmetry, the above results apply to the case of a
whole equatorial belt of crust fracturing at once; the width and
length of the fracture\footnote{In Cartesian coordinates, the strain
  plots of figure \ref{strains} are in the $x-z$ plane. Since the
  fractures we consider are centred around the equator, we use the
  term `depth' to refer to the size of the fracture in the
  $x$-direction, `length' to refer to the size across the surface of
  the star, i.e. in the $y-$direction, and `width' to refer to the
  fracture's size in the $z$-direction.} are therefore not
independent of the depth, and so we obtain relations in terms of
this one fracture dimension, instead of all three. Whilst the width
and depth of the fracture are both related to the crustal thickness,
the length $l$ is related to the larger scale of the
circumference of the star $2\pi R_*=20\pi R_\textrm{c}$. To reflect
this, we can modify equation \eqref{E-d_shallow} by replacing one factor
of $d/R_\textrm{c}$ with the term $l/2\pi R_*$ to reflect the expected relationship if the length of the
fracture does not extend right across the star:
\beq \label{E-dl}
\frac{E_{\textrm{quake}}}{10^{45}\textrm{ erg}}
  \approx 0.25\brac{\frac{\sigma_{\textrm{max}}}{0.001}}
                       \brac{\frac{d}{R_\textrm{c}}}^2
                       \brac{\frac{l}{2\pi R_*}}.
\eeq
Note that our results are only quantitatively correct for our
particular (axisymmetric) crust-yielding scenario though, so the above
relation is an approximate one. Now, from the definition of magnetic
energy as a volume integral of $B^2$,
we see that its dimensions are $[E]=[B]^2 L^3$; we can
therefore use our quake energy-depth relations to find a
characteristic field strength associated with the crust yielding. In
particular, if we take the shallow-fracture formula
\eqref{E-dl} and multiply through by $10^{45}$ erg and
$R_\textrm{c}^3=(1.2\times 10^5\textrm{ cm})^3$ we get
\beq \label{E-dl_dim}
E_\textrm{quake}\approx 2.3\times 10^{27}
                                 \brac{\frac{\sigma_\textrm{max}}{0.001}}
                                 d^2 l\textrm{ erg},
\eeq
where we have also used the fact that the ratio of fracture depth to
length $d/l\approx R_\textrm{c}/2\pi R_* = 1/20\pi$.
Equation \eqref{E-dl_dim} gives us a relation in physical units
between quake energy, depth and length, with a constant of
proportionality $2.3\times 10^{27}(\sigma_\textrm{max}/0.001)$, which
must therefore have dimensions of $[B]^2$. Given the expression for magnetic energy release
\eqref{E_quake}, we choose to define a characteristic field strength
$B_\textrm{break}$ for breaking a cubic region of crust by equating
the constant of proportionality from \eqref{E-dl_dim} with
$8\pi B_\textrm{break}^2$. From equation \eqref{E-dl_dim} this then
gives\footnote{Our argument uses an impure form of dimensional
  analysis, as we have included the factor of $8\pi$ from the magnetic
  energy expression and the $1/20\pi$ factor from the fracture
  depth-to-length ratio, since both factors are greater than an order
  of magnitude in themselves. Readers uncomfortable with the inclusion
  of these extra factors can remove them from the final result for
  $B_\textrm{break}$ by multiplying by $\sqrt{8\pi/20\pi}$,
    resulting in a prefactor of $3.8\times 10^{14}$ instead of the
    value of $2.4\times 10^{14}$ in equation \eqref{B_char}.}
\beq \label{B_char}
B_\textrm{break}=2.4\times 10^{14}\brac{\frac{\sigma_\textrm{max}}{0.001}}^{1/2}\textrm{ G}.
\eeq
We interpret this result to mean that, although the quake energy-depth
relation does not involve the field strength itself, there is
nonetheless a characteristic (\emph{local}) strength of field related to
crust-breaking.

\section{Discussion}
\label{discussion}

Neutron stars display a variety of abrupt energetic phenomena -- most
spectacularly the giant flares of magnetars, but also smaller
bursts, and glitches in their rotation rate. These phenomena all point
to some sudden release of stress that has built up gradually -- and
the star's elastic crust is a natural candidate for a region that can
become gradually stressed then fail suddenly. It is, therefore, worth
concluding with a discussion of the possible role of
magnetically-induced crustquakes in flares, bursts and glitches.

We turn first to a class of phenomena for which crustquakes have traditionally
\emph{not} been invoked: the giant flares of magnetars. The three
events observed to date have all involved energy outputs in excess of
$10^{44}$ erg \citep{fenimore,feroci,palmer}, an amount thought to be too great to have come from
crustal energy release alone \citep{thom_dunc95}; this worry, in part, has motivated a number of studies
exploring the alternative possibility that spontaneous reconnection in the magnetosphere is
responsible for magnetar flares, in analogy with dynamics in the solar corona (see, e.g.,
\citet{lyutikov03}). One key result of our paper is that a crust stressed by magnetic-field
rearrangement \emph{can}, in fact, comfortably store the required amount of energy to power
a giant flare.

The most energetic observed giant flare to date was the 2004
event of SGR 1806-20; its estimated energy output over the flare was
an enormous $2\times 10^{46}$ erg \citep{palmer}. This value is not
very precise -- in particular, the probable anisotropic nature of the
flare would make it an overestimate -- but let us nonetheless assume
that this amount of energy was released from a crustquake. From
equation \eqref{E-d_deep}, we then have an estimate
that the minimum breaking strain of the crust must be around $0.065$
(corresponding to the case of a fracture extending to the base of the
crust). This value is comfortably below the
recent result, obtained from molecular dynamics simulations, that a NS
crust has a breaking strain of $0.12$
\citep{horo_kad}. These simulations also show that
the crust fails in a large-scale collective fashion -- this could
conceivably fit the observed behaviour of giant flares, whose
luminosity peaks rapidly then decays exponentially \citep{palmer}.

We can also use
our results to put an upper limit on the expected maximum size of a
giant flare powered by crustal energy release alone. Taking a breaking
strain of $0.12$ and assuming a fracture extending to the base of the
crust, equation \eqref{E-d_deep} gives a maximum total energy release of
$4\times 10^{46}$ erg. If any future giant flare appears to be more
energetic than this (using the isotropic-emission assumption),
then either the energy release is not crustal in origin, or it is
highly anisotropic -- leaving current estimates for flare energies
seriously in error.

In addition to the rare giant flares, magnetars also suffer far more
common short-duration bursts with energies up to $\sim\! 10^{41}$
erg and intermediate events with energies around $10^{43}$ erg. If these bursts are also a manifestation of crustquakes, they
must involve the yielding of much more shallow regions. Unlike the
highly rigid inner regions of the crust, the outermost part of the
crust can only support small stresses, and could feasibly fail at
lower strains through some gradual process (like plastic flow or a
succession of small fractures) instead of one large collective failure; this would account for the groups of small bursts seen
from some sources \citep{maz99,mer08}. Assuming short bursts are indeed powered
by the release of crustal energy, equation \eqref{E-d_shallow}
suggests that a $10^{41}$-erg event would be associated with the magnetar's
crust yielding to a depth of roughly $90$m (for a breaking strain of $0.001$), or
to a depth of $20$m (if the breaking strain is $0.1$). Interestingly,
the burst afterglow of Swift J$1822.3-1606$ has been shown to be well
modelled by a $3\times 10^{42}$-erg shallow-depth heat deposition
\citep{scholz} -- which could have resulted from a magnetically-induced
crustquake; \red{see also \citet{rea2013} for similar outburst
  modelling for SGR $0418+5729$ and \citet{camero} for SGR $0501+4516$.} A period of
burst activity might indicate the gradual failure of a somewhat deeper
region; our energy-depth formulae should still be valid for this case,
but with the crustquake energy release being the \emph{total} energy
output over the period of bursting.

The final class of abrupt phenomena we wish to mention are
glitches. Unlike flares and bursts, these spin-up events cannot be due
to magnetically-induced crustquakes, since the resulting change in the
stellar moment of inertia due to such a crustquake event could only ever be
minute: it scales with the ratio of magnetic to fluid
pressure. Instead, we expect the usual glitch scenario to apply even for
highly-magnetised NSs: the star's superfluid component cannot
spin down regularly with the crust and so develops a difference in
angular velocity; beyond some critical value, however, the superfluid is
forced to re-equilibrate with the crust by transferring angular
momentum, which is then seen as a spin-up of the crust
\citep{and_itoh}. Nonetheless, it may not be safe to assume that the
magnetic field can be neglected in the treatment of glitch modelling.
As discussed in the introduction,
radiative changes associated with glitches have been observed in AXPs,
and moreover in at least three typically rotationally-powered pulsars
with high magnetic fields \citep{ant_1119}. These observations may point to magnetically-induced
crustquake activity occurring simultaneously -- either
as a trigger or a result of the glitch.

Finally, we have identified a characteristic field strength \eqref{B_char} associated with
crust-breaking, corresponding to the constant of proportionality in
the quake energy-depth relation. It suggests that for a crustquake to
occur, the field strength must reach approximately $10^{14}-10^{15}$ G
locally (depending on the crustal breaking
strain); this \red{is in agreement with the findings of
  \citet{pons_perna}, who considered a different scenario for
  the build-up of magnetically-induced stresses.}
Superficially, it appears as if this characteristic field strength
might only be attained in magnetars -- but in fact, the observed field
strengths of NSs are just inferences about the dipolar field component at the
polar cap. It is quite likely that NSs with inferred dipole fields of
the order of $10^{13}$ G, or perhaps lower still, will harbour some
region in their crust where the local field exceeds $10^{14}$
G. Within our crustquake model, therefore, it would be quite
natural to find crossover sources displaying both `radio-pulsar' and
`magnetar' characteristics -- and we anticipate the distinctions
between supposedly different classes of NS to become further eroded over time.

\section*{Acknowledgements}

SKL, DA and ALW acknowledge support from NWO Vidi and Aspasia grants (PI Watts);
NA acknowledges support from STFC in the UK. We thank
Christian Kr\"uger for discussions on modelling the
neutron-star crust, Chuck Horowitz for helpful correspondence about
the breaking strain, and Jose Pons and the anonymous referee for their
constructive suggestions.

\label{lastpage}


\small
\begin{thebibliography}{99}
\bibitem[\protect\citeauthoryear{Akg\"un \& Wasserman}{2008}]{akgun_wass}
  Akg\"un T., Wasserman I., 2008, MNRAS 383, 1551
\bibitem[\protect\citeauthoryear{Alpar et al.}{1994}]{accp94}
  Alpar M.A., Chau H.F., Cheng K.S., Pines D., 1994, ApJ 427, L29
\bibitem[\protect\citeauthoryear{Anderson \& Itoh}{1975}]{and_itoh}
  Anderson P.W., Itoh N., 1975, Nature 256, 25
\bibitem[\protect\citeauthoryear{Antonopoulou et al.}{2015}]{ant_1119}
  Antonopoulou D., Weltevrede P., Espinoza C.M., Watts A.L., Johnston
  S., Shannon R.M., Kerr M., 2015, MNRAS 447, 3924
\bibitem[\protect\citeauthoryear{Baym, Pethick \& Pines}{1969a}]{baym_pp}
  Baym G., Pethick C., Pines D., 1969, Nature 224, 673
\bibitem[\protect\citeauthoryear{Baym, Pethick \& Pines}{1969b}]{baym_pp2}
  Baym G., Pethick C., Pines D., 1969, Nature 224, 674
\bibitem[\protect\citeauthoryear{Baym et al.}{1969}]{baym_ppr}
  Baym G., Pethick C., Pines D., Ruderman M., 1969, Nature 224, 872
\bibitem[\protect\citeauthoryear{Beloborodov}{2009}]{belo09}
  Beloborodov A.M., 2009, ApJ 703, 1044 
\bibitem[\protect\citeauthoryear{Beloborodov \& Levin}{2014}]{belo_lev}
  Beloborodov A.M., Levin Y., 2014, ApJ 794, L24 
\bibitem[\protect\citeauthoryear{Beloborodov \& Thompson}{2007}]{belo_thomp}
  Beloborodov A.M., Thompson C., 2007, ApJ 657, 967 
\bibitem[\protect\citeauthoryear{Braithwaite \& Spruit}{2006}]{braith_spruit}
  Braithwaite J., Spruit H.C., 2006, A\&A 450, 1097 
\bibitem[\protect\citeauthoryear{Camero et al.}{2014}]{camero}  
  Camero A., Papitto A., Rea N., Vigan\`o D., Pons J.A., Tiengo A.,
  Mereghetti S., Turolla R. et al, 2014, MNRAS 438, 3291
\bibitem[\protect\citeauthoryear{Cheng et al.}{1996}]{cheng96}  
  Cheng B., Epstein R.I., Guyer R.A., Young A.C., 1996, Nature 382, 518
\bibitem[\protect\citeauthoryear{Chugunov \& Horowitz}{2010}]{chug_horo}
  Chugunov A.I., Horowitz C.J., MNRAS 407, L54
\bibitem[\protect\citeauthoryear{Dib \& Kaspi}{2014}]{dib_kaspi14}
  Dib R., Kaspi V.M., 2014, ApJ 784, 37
\bibitem[\protect\citeauthoryear{Ding, Cheng \& Chau}{1993}]{ding_cc}
  Ding K.Y., Cheng K.S., Chau H.F., 1993, ApJ 408, 167
\bibitem[\protect\citeauthoryear{Douchin \& Haensel}{2001}]{douchin}
  Douchin F., Haensel P., 2001, A\&A 380, 151
\bibitem[\protect\citeauthoryear{Easson \& Pethick}{1977}]{easson_peth}
  Easson I., Pethick C. J., 1977, Phys. Rev. D 16, 275 
\bibitem[\protect\citeauthoryear{Eichler \& Shaisultanov}{2010}]{Eich10}
  Eichler D., Shaisultanov R., 2010, ApJ 715, L142
\bibitem[\protect\citeauthoryear{Fenimore, Klebesadel \& Laros}{1996}]{fenimore}
  Fenimore E.E., Klebesadel R.W., Laros J.G., 1996, ApJ 460, 964
\bibitem[\protect\citeauthoryear{Feroci et al.}{2001}]{feroci}
  Feroci M., Hurley K., Duncan R.C., Thompson C., 2001, ApJ 549, 1021
\bibitem[\protect\citeauthoryear{Franco, Link \& Epstein}{2000}]{fle00}
  Franco L.M., Link B., Epstein R.I., 2000, ApJ 543, 987
\bibitem[\protect\citeauthoryear{Gavriil et al.}{2008}]{ggg+08}
  Gavriil F.P., Gonzalez M.E., Gotthelf E.V., Kaspi V.M., Livingstone
  M.A., Woods P.M., 2008, Science 319, 1802
\bibitem[\protect\citeauthoryear{Glampedakis, Andersson \& Samuelsson}{2011}]{GAS}
  Glampedakis K., Andersson N., Samuelsson L., 2011, MNRAS 410, 805 
\bibitem[\protect\citeauthoryear{Glampedakis, Jones \& Samuelsson}{2011}]{GJS}
  Glampedakis K., Jones D.I., Samuelsson L., 2011, MNRAS 413, 2021
\bibitem[\protect\citeauthoryear{Glampedakis, Lander \& Andersson}{2014}]{GLA}
  Glampedakis K., Lander S.K., Andersson N., 2014, MNRAS 437, 2
\bibitem[\protect\citeauthoryear{Gnedin, Yakovlev \& Potekhin}{2001}]{gnedin}
  Gnedin O.Y., Yakovlev D.G., Potekhin A.Y., 2001, MNRAS 324, 725
\bibitem[\protect\citeauthoryear{Goldreich \& Reisenegger}{1992}]{gold_reis}
  Goldreich P., Reisenegger A., 1992, ApJ 395, 250
\bibitem[\protect\citeauthoryear{G{\"o}{\v g}{\"u}{\c s} et al.}{2000}]{gogus00}
  G{\"o}{\v g}{\"u}{\c s} E., Woods P.M., Kouveliotou C., van Paradijs J., Briggs M.S., Duncan R.C., Thompson C., 2000, ApJ 532, L121
\bibitem[\protect\citeauthoryear{Harrison}{1991}]{harrison91}
  Harrison E., 1991, MNRAS 248, 419
\bibitem[\protect\citeauthoryear{Hoffman \& Heyl}{2012}]{hoffman_heyl}
  Hoffman K., Heyl J.S., 2012, MNRAS 426, 2404
\bibitem[\protect\citeauthoryear{Horowitz}{2015}]{horo_private}
  Horowitz C.J., 2015, private communication
\bibitem[\protect\citeauthoryear{Horowitz \& Hughto}{2008}]{horo_hugh}
  Horowitz C.J., Hughto J., 2008, arXiv:0812.2650
\bibitem[\protect\citeauthoryear{Horowitz \& Kadau}{2009}]{horo_kad}
  Horowitz C.J., Kadau K., 2009, Phys. Rev. Lett. 102, 191102
\bibitem[\protect\citeauthoryear{Ioka}{2001}]{ioka}
  Ioka K., 2001, MNRAS 327, 639 
\bibitem[\protect\citeauthoryear{Jones}{1991}]{jones91}
  Jones P.B., 1991, MNRAS 253, 279 
\bibitem[\protect\citeauthoryear{Jones}{2003}]{jones03}
  Jones P.B., 2003, ApJ 595, 342
\bibitem[\protect\citeauthoryear{Kaminker et al.}{2009}]{kaminker_2009}
  Kaminker A.D., Potekhin A.Y., Yakovlev D.G., Chabrier G., 2009, MNRAS 395, 2257
\bibitem[\protect\citeauthoryear{Kaspi}{2010}]{kaspi}
  Kaspi V., 2010, P.N.A.S. 107, 7147
\bibitem[\protect\citeauthoryear{Kocharovsky, Kocharovsky \& Kukushkin}{1996}]{kocharovsky}
  Kocharovsky V.V., Kocharovsky Vl.V., Kukushkin V.A., 1996,
  Radiophys. and Quant. Electronics 39, 18
\bibitem[\protect\citeauthoryear{Kr\"uger, Ho \& Andersson}{2014}]{krueger}
  Kr\"uger C.J., Ho W.C.G., Andersson N., 2014, arxiv:1402.5656
\bibitem[\protect\citeauthoryear{Kuiper \& Hermsen}{2009}]{kh09}
  Kuiper L., Hermsen W., 2009, A\&A 501, 1031
\bibitem[\protect\citeauthoryear{Lander, Andersson \& Glampedakis}{2012}]{LAG}
  Lander S.K., Andersson N., Glampedakis K., 2012, MNRAS 419, 732
\bibitem[\protect\citeauthoryear{Lander}{2013}]{sc_eqm_letter}
  Lander S.K., 2013, Phys. Rev. Lett. 110, 071101
\bibitem[\protect\citeauthoryear{Lander}{2014}]{sc_eqm_paper}
  Lander S.K., 2014, MNRAS 437, 424
\bibitem[\protect\citeauthoryear{Levin \& Lyutikov}{2012}]{lev_lyu}
  Levin Y., Lyutikov M., 2012, MNRAS 427, 1574 
\bibitem[\protect\citeauthoryear{Link \& Epstein}{1996}]{le96}
  Link B., Epstein R.I., 1996, ApJ 457, 844
\bibitem[\protect\citeauthoryear{Livingstone, Kaspi \& Gavriil}{2010}]{lkg10}
  Livingstone M.A., Kaspi V.M., Gavriil F.P., 2010, ApJ 710, 1710
\bibitem[\protect\citeauthoryear{Livingstone et al.}{2011}]{lnk+11}
  Livingstone M.A., Ng C.-Y., Kaspi V.M., Gavriil F.P., Gotthelf E.V.,
  2011, ApJ 730, 66
\bibitem[\protect\citeauthoryear{Lyutikov}{2003}]{lyutikov03}
  Lyutikov M., 2003, MNRAS 346, 540
\bibitem[\protect\citeauthoryear{Mazets et al.}{1999}]{maz99}  
Mazets E.P., Aptekar R.L., Butterworth P.S., Cline T.L., Frederiks D.D., Golenetskii S.V., Hurley K., Il'inskii V.N., 1999, ApJ 519, L151  
\bibitem[\protect\citeauthoryear{Melatos, Peralta \& Wyithe}{2008}]{mpw08}
  Melatos A., Peralta C., Wyithe J.S.B., 2008, ApJ 672, 1103
\bibitem[\protect\citeauthoryear{Mereghetti et al.}{2009}]{mer08}
Mereghetti S., {G{\"o}tz} D., Weidenspointner G., von Kienlin A.,
Esposito P., Tiengo A., Vianello G., Israel G.L. et al., 2009, ApJ 696, L74
\bibitem[\protect\citeauthoryear{Muslimov \& Tsygan}{1985}]{mus_tsyg}
  Muslimov A.G., Tsygan A.I., 1985, Sov. Astron. Lett. 11, 80
\bibitem[\protect\citeauthoryear{Palmer et al.}{2005}]{palmer}
  Palmer D.M., Barthelmy S., Gehrels N., Kippen R.M., Cayton T.,
  Kouveliotou C., Eichler, D., Wijers, R.A.M.J. et al., 2005, Nature 434, 1107
\bibitem[\protect\citeauthoryear{Payne \& Melatos}{2004}]{payne_mel}
  Payne D.J.B., Melatos A., 2004, MNRAS 351, 569
\bibitem[\protect\citeauthoryear{Perna \& Pons}{2011}]{perna_pons}
  Perna R., Pons J.A., 2011, ApJ 727, L51
\bibitem[\protect\citeauthoryear{Pons, Miralles \& Geppert}{2009}]{pons_mu}
  Pons J.A., Miralles J.A., Geppert U., 2009, A\&A 496, 207 
\bibitem[\protect\citeauthoryear{Pons \& Perna}{2011}]{pons_perna}
  Pons J.A., Perna R., 2011, ApJ 741, 123
\bibitem[\protect\citeauthoryear{Pons \& Rea}{2012}]{PonsRea12}
  Pons J.A., Rea N., 2012, ApJ 750, L6
\bibitem[\protect\citeauthoryear{Prix, Novak \& Comer}{2005}]{prix_novcom}
  Prix R., Novak J., Comer G.L., 2005, PRD 71, 043005
\bibitem[\protect\citeauthoryear{Ravenhall, Pethick \& Wilson}{1983}]{ravenhall}
  Ravenhall D.G., Pethick C.J., Wilson J.R., 1983,
  Phys. Rev. Lett. 50, 26 
\bibitem[\protect\citeauthoryear{Rea et al.}{2010}]{rea10}
  Rea N., Esposito P., Turolla R., Israel G.L., Zane S., Stella L.,
  Mereghetti S., Tiengo A. et al., 2010, Science 330, 944 
\bibitem[\protect\citeauthoryear{Rea et al.}{2013}]{rea2013}
  Rea N., Israel G.L., Pons J.A., Turolla R., Vigan\`o D., Zane S.,
  Esposito P., Perna R. et al., 2013, ApJ 770, 65 
\bibitem[\protect\citeauthoryear{Ruderman}{1968}]{ruderman_1968}
  Ruderman M., 1968, Nature 218, 1123 
\bibitem[\protect\citeauthoryear{Ruderman}{1969}]{ruderman_1969}
  Ruderman M., 1969, Nature 223, 597 
\bibitem[\protect\citeauthoryear{Ruderman, Zhu \& Chen}{1998}]{rzc98}
  Ruderman M., Zhu T., Chen K., 1998, ApJ 492, 267
\bibitem[\protect\citeauthoryear{Sinha \& Sedrakian}{2014}]{sin_sed}
  Sinha M., Sedrakian A., 2014, arXiv:1403.2829
\bibitem[\protect\citeauthoryear{Scholz et al.}{2012}]{scholz}
  Scholz P., Ng C.-Y., Livingstone M.A., Kaspi V.M., Cumming A.,
  Archibald R.F., 2012, ApJ 761, 66
\bibitem[\protect\citeauthoryear{Tayler}{1980}]{tayler_mix}
  Tayler R.J., 1980, MNRAS 191, 151
\bibitem[\protect\citeauthoryear{Tendulkar, Cameron \& Kulkarni}{2012}]{tendulkar}
  Tendulkar S.P., Cameron P.B., Kulkarni S.R., 2012, ApJ 761, 76 
\bibitem[\protect\citeauthoryear{Thompson \& Duncan}{1995}]{thom_dunc95}
  Thompson C., Duncan R.C., 1995, ApJ 275, 255 
\bibitem[\protect\citeauthoryear{Thompson \& Duncan}{1996}]{thom_dunc96}
  Thompson C., Duncan R.C., 1996, ApJ 473, 322
\bibitem[\protect\citeauthoryear{Wendell}{1988}]{wendell}
  Wendell C.E., 1988, ApJ 333, L95
\bibitem[\protect\citeauthoryear{Woods \& Thompson}{2006}]{wt06}
  Woods P.M., Thompson C., 2006, \emph{Soft gamma repeaters and
    anomalous X-ray pulsars: magnetar candidates} in Compact stellar
  X-ray sources, ed. W.Lewin \& M. van der Klis, Cambridge University Press  
\bibitem[\protect\citeauthoryear{Yoshida, Yoshida \& Eriguchi}{2006}]{yosh_yosh_eri}
  Yoshida S., Yoshida S., Eriguchi Y., 2006, ApJ 651, 462

\end{thebibliography}
\normalsize
\end{document}